\documentclass[a4paper,11pt]{llncs}
\usepackage{amssymb,amsmath}
\usepackage{afterpage,epsfig,float,color}
\usepackage{citesort,natbib}

\newcommand{\pr}{\text{Pr}}
\newcommand{\aic}{\text{AIC}}

\newcommand{\lk}{\text{lk}}
\newcommand{\ie}{\it i.e. \rm}
\newcommand{\eg}{\it e.g. \rm }

\newcommand{\lb}[1]{\raisebox{-2.ex}[2.ex]{#1}}
\begin{document}
\title{Statistical model selection methods applied to biological networks} 
\author{Michael P.H. Stumpf$^{1*}$ \and Piers J. Ingram$^1$ \and Ian Nouvel$^1$ \and Carsten Wiuf$^2$}
\institute{$^1$Centre for Bioinformatics, Department of Biological Sciences,Wolfson Building, Imperial College London, London SW7 2AZ, UK,\\
$^2$ Bioinformatics Research Center, University of Aarhus, 8000 Aarhus C, Denmark\\
$^*$ corresponding author: \email{m.stumpf@imperial.ac.uk}}

\maketitle
\begin{abstract}
Many biological networks have been labelled scale-free as their degree distribution can be approximately described by a powerlaw distribution. While the degree distribution does not summarize all aspects of a network it has often been suggested that its functional form contains important clues as to underlying evolutionary processes that have shaped the network. Generally determining the appropriate functional form for the degree distribution has been fitted in an ad-hoc fashion. 

Here we apply formal statistical model selection methods to determine which functional form best describes degree distributions of protein interaction and metabolic networks. We interpret the degree distribution as belonging to a class of probability models and determine which of these models provides the best description for the empirical data using maximum likelihood inference, composite likelihood methods, the Akaike information criterion and goodness-of-fit tests. The whole data is used in order to determine the parameter that best explains the data under a given model (\eg scale-free or random graph).   As we will show, present protein interaction and metabolic network data from different organisms suggests that simple scale-free models do not provide an adequate description of real network data.
\end{abstract}

\section{Introduction}
Network structures which connect interacting particles such as proteins have long been recognised to be linked to the underlying  dynamic or evolutionary processes\cite{dor_book,albert_RMP2002}. In particular the technological advances seen in molecular biology and genetics increasingly provide us with vast amounts of data about genomic, proteomic and metabolomic network structures \cite{ito_PNAS2000,wagner_MBE2001,qin_PNAS2003}. Understanding the way in which the different constituents of such networks, --- genes and their protein products in the case of genome regulatory networks, enzymes and metabolites in the case of metabolic networks (MN), and proteins in the case of protein interaction networks (PIN) --- interact can yield important insights into basic biological mechanisms \cite{maslov_S2002,yook_proteomics2004,Agrafioti2005}. For example the extent of phenotypic plasticity allowed for by a network, or levels of similarity between PINs in different organisms presumably depend on topological (in a loose sense of the word) properties of networks.
\par
Our analysis here focuses on the degree distribution of a network, \ie the probability of a node to have $k$ connections to other nodes in the network. While it is well known that this does not offer an exhaustive description of network data, it has nevertheless remained an important characteristic/summary statistic of network data. Here we use $\pr(k)$ to denote a theoretical model for the degree distribution, or $\pr(k;\theta)$ if the model depends on an (unknown) parameter $\theta$ (potentially vector-valued), and $\hat{\pr}(k)$ to denote the empirical degree distribution.
\par
Many studies of biological network data have suggested that the underlying networks show scale-free behaviour \cite{bollobas_handbook} and that their degree distributions follow a power-law, \ie
\begin{equation}
\pr(k;\gamma) = k^{-\gamma}/\zeta(\gamma)
\end{equation}
 where $\zeta(x)$ is Riemann's zeta-functions which is defined for $x>1$ and diverges as $x\rightarrow 1\downarrow$; for finite networks, however, it is not necessary that the value of $\gamma$ is restricted to values greater than 1. 
\par
These powerlaws are in marked contrast to the degree distribution of the Erd\"os-R\'enyi random graphs \cite{bollobas_book1} which is Poisson, $\pr(k;\lambda)=e^{-\lambda}\lambda^k/k!$. The study of  random graphs is a rich field of research and many important properties can be evaluated analytically. Such Poisson random networks (PRN) are characterized by most nodes having comparable degree; the vast majority of nodes will have a connectivity close to the average connectivity.
\par
The term "scale-free" means that the ratio $\pr(\alpha k)/\pr(k)$ depends on $\alpha$ alone but not on the connectivity $k$. The attraction of scale-free models stem from the fact that some simple and intuitive statistical models of network evolution cause powerlaw degree distribution. Scale-free networks are not, however, the only type of network that produces fat-tailed degree distributions. 
\par
Here we will be concerned with developing a statistically sound approach for inferring the functional form for the degree distribution of a real network. We will show that relatively basic statistical concepts, like maximum likelihood estimation and model selection can be straightforwardly applied to PIN and MN data. In particular we will demonstrate how we can determine which probability models best describe the degree distribution of a network. We then apply this approach in the analysis of real PIN data from five model organisms and MN data. In each case we can show that the explanatory power of a standard scale-free network is vastly inferior compared to models that take the finite size of the system into account.
 
\section{Statistical tools for the analysis of network data}
 Here we are only concerned with methods aimed at studying the degree distribution of a network. In particular we want to quantify the extent to which a given functional form can describe the degree distribution of a real network. Given a probability model (\eg power-law distribution or Poisson distribution) we want to determine the parameters which describe the degree distribution best; after that we want to be able to distinguish which model from a set of trial model provides the best description. Here we briefly introduce the basic statistical concepts employed later. These can be found in much greater detail in most modern statistics texts such as \cite{davison_book1}. Tools for the analysis of other aspects of network data, \eg cluster coefficients, path length or spectral properties of the adjacency matrix will also need to be developed in order to understand topological and functional properties of networks.
\par
 There is a well established statistical literature that allows us to assess to what extent data (\eg the degree distribution of a network) is described by a specific probability model (\eg Poisson, exponential or powerlaw distributions). Thus far, determining the best model appears to have been done largely by eye \cite{dor_book} and it is interesting to apply a more rigorous approach, although in some published cases maximum likelihood estimates were used to determine the value of $\gamma$ for the scale-free distribution.

\subsection{Maximum likelihood inference}

\begin{table*}
\begin{center}
\begin{tabular}{|l|cl|c|}
\hline
{\bf Network type}&\multicolumn{2}{c|}{\bf Degree distribution $\pr(k;\theta)$}&{\bf Model}\\
\hline
Poisson &$\exp(-\lambda)\frac{\lambda^k}{k!}$& for all $k\ge 0$&{\bf M1}\\
\hline
Exponential&$ C\exp(-k/\bar{k})$& for all $k\ge 0$&{\bf M2}\\
\hline
Gamma&$\frac{k^{\gamma-1}e^{-k}}{\Gamma(\gamma)}$& for all $k\ge 0$&{\bf M3}\\
\hline
\lb{Scale-free} &$0 $& for $k=0$&\lb{\bf M4}\\
&$  k^{-\gamma}/\zeta(\gamma)$& for $k>0$&\\
\hline
\lb{Truncated scale free network} &$ 0$& for $k<L$ and $k>M$&\lb{\bf M4a}\\
&$ k^{-\gamma}/\sum_{i=L}^Mk^{-\gamma}$& for $L\le k\le M$&\\
\hline
Scale-free network&$ 0$& for $k<k_0$ and $k>k_{\text{cut}}$&\lb{\bf M4b}\\
with exponential cut-off&$(k+k_0)^{-\gamma}\exp(-k/k_{\text{cut}})$& for $k_0\le k \le k_{\text{cut}}$&\\
\hline

Lognormal&$C \frac{e^{-\ln((k-\theta)/m)^2/(2\sigma^2)}}{(k-\theta)\sigma\sqrt{2\pi}}$& for all $k\ge 0$&{\bf M5}\\
\hline
\lb{Stretched exponential} &$ 0$& for $k<0$& \lb{\bf M6}\\
& $C\exp(-\alpha k/\bar{k}) k^{-\gamma}$ &for $k>0$&\\
\hline
\end{tabular}
\vskip3mm
\caption{Network models and their degree distributions, $\pr(k;\theta)$. Wherever it appears, $C$ denotes the normalizing constant such that $\sum_{k}\pr(k;\theta)=1$. }

\label{table1}
\end{center}
\end{table*}
Since we only specify the marginal probability distribution, \ie the degree distribution, we take a
composite likelihood approach to inference, and treat the degrees of nodes as independent observations.
This is only correct in the limit of an infinite sized network and finite sized sample ($n<<N$, where $N$ denotes
network size and $n$ the sample size). Composite likelihood methods are becoming increasingly
popular in cases for which the full likelihood is difficult to specify and/or the full likelihood
is intractable to calculate numerically. In our case the full likelihood is difficult to specify. 
Reference \cite{cox_reid} provides an overview of composite likelihood methods. 
\par
For a given 
 functional form or model $\pr(k;\theta)$ of the degree distribution we can use maximum  likelihood estimation applied to the composite likelihood in order to estimate the parameter which best characterizes the distribution of the data.
The composite likelihood of the model given the observed data $K=\{k_1,k_2,\ldots,k_n\}$ is defined by 
\begin{equation} 
 L(\theta) =\prod_{i=1}^n \pr(k_i;\theta),
\label{likelihood}
\end{equation}
and taking the logarithm yields the log-likelihood
\begin{equation}
\lk(M) = \lk(\theta) = \sum_{i=1}^n\log(\pr(k_i;\theta)).
\label{loglikelihood}
\end{equation}
\par
 The maximum likelihood estimate (MLE), $\hat{\theta}$, of $\theta$ is the value of $\theta$ for which Eqns. (\ref{likelihood}) and (\ref{loglikelihood}) become maximal. For this value the observed data is more probable to occur than for any other parameters. 
\par
Here the maximum likelihood framework is applied to the whole of the data. This means that in fitting a curve ---such as a powerlaw $k^{-\hat{\gamma}}/\zeta(\hat{\gamma})$, where $\hat{\gamma}$ denotes the MLE of the exponent $\gamma$--- data for all $k$ is considered. If a powerlaw-dependence where to exist only over a limited range of connectivities then the global MLE curve may differ from such a localized power-law (or equivalently any other distribution). 

\subsection{Model selection and Akaike weights}  
We are interested in determining which model describes the data best. For non-nested models (as are considered here, \eg scale-free versus Poisson) we cannot use the standard likelihood ratio test but have to employ a different information criterion to distinguish between models: here we use the Akaike-information criterion (AIC) to choose between  different models \cite{akaike_proc1983,burnham_book}. The AIC for a model $\pr(k;\theta)$ is defined by
\begin{equation}
 \aic = 2(-\lk(\hat{\theta})+d),
\end{equation}
 where $\hat{\theta}$ is the maximum liklihood estimate of $\theta$ and $d$ is the number of parameters required to define the model, \ie the dimension of $\theta$. Note that the model is penalized by $d$.
The model with the minimum AIC is chosen as the best model and the AIC therefore formally biases against overly complicated models.  A more complicated model is only accepted as better if it contains more information about the data than a simpler model. (It is possible to formally derive the AIC from Kohn-Sham information theory.) Other information criteria exist, \eg the Bayesian information criterion (BIC) offers a further method for penalizing more complex models (\ie those with more parameters) unless they have significantly higher explanatory power (see \cite{burnham_book} for details about the AIC and model selection in statistical inference). In order to compare different models we define the relative differences 
\begin{equation}
\Delta^{\aic}_j = \aic_j-\min_j(\aic),
\end{equation}
where $j$ refers to the $j$th model, $j=1,2,\ldots,J$, and $\min_j$ is minimum over all $j$.This in turn allows us to calculate the relative likelihoods (adjusted for the dimension
of $\theta$) of the different models, given by 
\begin{equation}
\exp(-\Delta^{\aic}_j/2).
\end{equation}
Normalizing these relative likelihoods yields the so-called Akaike weights $w_j$,
\begin{equation}  
w_j =\frac{\exp(-\Delta^{\aic}_j/2)}{\sum_{j=1}^J\exp(-\Delta^{\aic}_j/2)}.
\end{equation}
 The Akaike weight $w_j$ can be interpreted as the probability that model $j$ (out of the $J$ alternative models) is the best model given the observed data and the range of models to choose from. The relative support for one model over another is thus given by the ratio of their respective Akaike weights. If a new model is added to the $J$ existing models then the analysis has to be repeated. The Akaike weight formalism is very flexible and has been applied in a range of context including the assessment of confidence in phylogenetic inference \cite{strimmer_proysoc2002}. In the next section we will apply this formalism to PIN data from five species and estimate the level of support for each of the models in table 1. 
\subsection{Goodness-of-fit}
 In addition to the AIC or similar information criteria we can also assess a model's performance at describing the degree distribution using a range of other statistical measures. The Kolmogorov-Smirnoff (KS)\cite{davison_book1} and Anderson-Darling (AD) \cite{Anderson1952,Anderson1954} goodness-of-fit statistics allow us to quantify the extent to which a theoretical or estimated model of the degree distribution describes the observed data. The former is a common and easily implemented statistic, but the latter puts more weight on the tails of distributions and also allows for a  secular dependence of the variance of the observed data on the argument (here the connectivity $k$). They KS statistic is defined as
\begin{equation}
D = \max |\hat{P}(k)- P(k)|,
\end{equation}
where $\hat{P}(k)$ and $P(k)$ are the empirical and theoretical cumulative distribution functions, respectively, for a node's degree \ie $P(k) =\sum_{i=1}^k \pr(i)$ and $\hat{P}(k) =\sum_{i=1}^k \hat{\pr}(i)$. If $P(k)$ depends on $\theta$, $P(k)$ is substituted by $P(k;\hat{\theta})=\sum_{i=1}^k \pr(i;\hat{\theta})$, the estimated cumulative distribution. This statistic is most sensitive to differences between the theoretical (or estimated) and observed distributions around the median of the data, \ie the point where $P(k)\approx0.5$. Given that we will also be considering a number of fat-tailed distributions this is somewhat unsatisfactory and we will therefore also use the AD statistic (Anderson and Darling discussed a number of statistics \cite{Anderson1952,Anderson1954}) which is defined as
\begin{equation}
D^* = \max\frac{|\hat{P}(k)-P(k)|}{\sqrt{P(k)(1-P(k))}}
\end{equation}
(again $P(k)$ might be substituted for $P(k;\hat{\theta})$).
\par
 We can use these statistics for two purposes: first, we can use them to compare different trial distributions as the "best" distribution should have the smallest value of $D$ and $D^*$, respectively. Second, we can use these statistics to determine if the empirical degree distribution is consistent with a given  theoretical (or estimated) distribution.

\par
 To evaluate the fit of a model, $p$-values can be calculated for the observed values of
$D$ and $D^*$ using a  parametric boot-strap procedure using the estimated degree distribution: for a network with $N$ nodes we repeatedly sample $N$ values at random from the maximum likelihood model, $\pr(k,\hat{\theta})$ and calculate $D^*$ and $D^*$, respectively, for each replicate. From  $L$ bootstrap-replicates we obtain the Null distribution of $D$ and $D^*$ under the estimated degree distribution. For sufficiently large $L$ we can thus determine approximate $p$-values which allow us to test if the empirical degree distribution is commensurate with the estimated degree distribution.

\section{Statistical analysis of biological networks}
Here we apply the analysis of the preceeding sections to the study of PINs and metabolic networks. It is easy to find a straight-line fit to some degree interval for all of the datasets considered here. For a powerlaw to be meaningful it has to extend over at least two or three decades and with a maximum degree of $k_{\max}\lesssim 300$ this will be unachievable for the present data sets.  We therefore use all the data and fit the model which yields the best overall  a description of the degree distribution. 
 
\subsection{Analysis of PIN data}
In table 2 we show the maximum composite likelihoods for the degree distributions calculated from PIN data collected in five model organisms \cite{xenarios_NAR2000} (the protein interaction data was taken from the DIP data-base; \url{http://dip.doe-mbi.ucla.edu}). We find that the standard scale-free model (or its finite size versions) never provides the best fit to the data; in three networks ({\it C.elegans, S.cerevisiae} and {\it E.coli}) the lognormal distribution (M5) explains the data best. In the remaining two organisms the stretched exponential model provides the best fit to the data. The bold likelihoods correspond to the highest Akaike weights. Apart from the case of {\it H.pylori} (where $\max(w_j)=w_5\approx 0.95$ for M5 and $w_6\approx 0.05$) the value of the maximum Akaike weight is always $>0.9999$. For {\it C elegans}, however, the scale-free model and its finite size versions are better than the lognormal model, M5.

\begin{table}[H]
\centering
\begin{tabular}{|c||c|c|c|c|c|c|c|c|}\hline
{\bf Organism}&{\bf M1}&{\bf M2}&{\bf M3}&{\bf M4}&{\bf M4a}&{\bf M4b}&{\bf M5}&{\bf M6}\\\hline
{\it D.melanogaster}&-38273&-20224&-29965&-18517&-18257&-18126&-17835&{\bf -17820}\\
{\it C.elegans}&-9017&-5975&-6071&-4267&-4258&-4252&-4328&{\bf -4248}\\
{\it S.cerevisiae}&-24978&-14042&-20342&-13454&-13281&-13176&{\bf -12713}&-12759\\
{\it H. pylori}&-2552&-1776&-2052&-1595&-1559&-1546&{\bf -1527}&-1529\\
{\it E.coli}&-834&-884&-698&-799&-789&-779&{\bf -659}&-701\\
\hline
\end{tabular}
\vskip1mm
\caption{Log-likelihoods for the degree distributions from table 1 in five model organisms. M1, M2, M3 and M4 have one free parameter, M4a, M4b and M5 have two free parameters, while M6 has three free parameters. The differences in the log-likelihoods are, however, so pronounced that the different numbers of parameters do not noticeably influence the AIC.  } 
\end{table}
\par
\afterpage{
\begin{figure}[H]
 \begin{center}
\epsfig{file=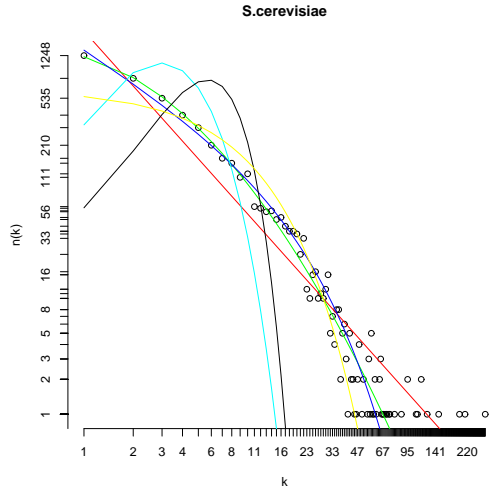,height=7cm}
\caption{Yeast protein interaction data (o) and best-fit probability distributions: Poisson (\textcolor{black}{\bf ---}), Exponential (\textcolor{yellow}{\bf ---}), Gamma (\textcolor{cyan}{\bf ---}), Power-law (\textcolor{red}{\bf ---}), Lognormal (\textcolor{green}{\bf ---}), Stretched exponential (\textcolor{blue}{\bf ---}). The parameters of the distributions shown in this figure are the maximum likelihood estimates based on the real observed data. }
\end{center}\end{figure}}

\par
For the yeast PIN the best fit curves (obtained from the MLEs of the parameters of models M1-M6) are shown in figure 1, together with the real data. Visually, log-normal (green) and stretched exponential (blue) appear to describe the date almost equally well. Closer inspection, guided by the Akaike weights, however, shows that the fit of the lognormal to the data is in fact markedly better than the fit of the stretched exponential. But the failure of quickly decaying distributions such as the Poisson distribution, characteristic for classical random graphs \cite{bollobas_book1} to capture the behaviour of the PIN degree distribution is obvious. 
\par
Interestingly, common heuristic finite size corrections to the standard scale-free model improve the fit to the data (measured by the AIC). But compared to the lognormal and stretched exponential models they still fall short in describing the PIN data in the five organisms. 
\begin{table}
\centering
\begin{tabular}{|c||c|c||c|c||c|c|}
\hline
{\bf Species}&\multicolumn{2}{c||}{\bf M4}&\multicolumn{2}{|c||}{\bf M5}&\multicolumn{2}{c|}{\bf M6}\\\cline{2-7}
&$D$&$D^*$&$D$&$D^*$&$D$&$D^*$\\
\hline
{\it D.melanogaster}&0.13&0.26&0.01&0.06&0.02&0.06\\
{\it C.elegans}&0.03&0.09&0.10&0.20&0.02&0.08\\
{\it S.cerevisiae}&0.17&0.33&0.01&0.04&0.03&5.99\\
{\it H. pylori}&0.13&0.26&0.01&0.05&0.02&0.12\\
{\it E.coli}&0.28&0.56&0.04&56.09&0.12&6072\\
\hline
\end{tabular}
\caption{The KS and AD statistics $D$ and $D^*$ for the scale-free, lognormal and stretched exponential model. The ordering of these models is in agreement with the AIC, but KS and AD statistics capture different aspects of the degree distribution. We see that the likelihood treatment, which takes in all the data, agrees better with the KS statistic. In the tails (especially for large connectivities $k$) the maximum likelihood fit sometimes ---especially in the case of {\it E.coli}--- can provide a very bad description of the observed data.}
\label{gof}
\end{table}
Figure 2 shows only the three curves with the highest values of $\omega_j$, which apart from {\it E.coli} are the log-normal, stretched exponential and power-law distributions; for {\it E.coli}, however, the Gamma distribution replaces the power-law distribution. These figures show that, apart from {\it C.elegans} the shape of the whole degree distribution is not power-law like, or scale-free like,  in a strict sense. Again we find that log-normal and stretched exponential distributions are hard to distinguish based on visual assessment alone. Figures 1 and 2, together with the results of table 2, reinforce the well known point that it is hard to choose the best fitting function based on visual inspection. It is perhaps worth noting, that the PIN data is more complete for {\it S.cerevisiae} and {\it D.melanogaster} than for the other organisms.

\afterpage{\begin{figure}
\begin{center}
\epsfig{file=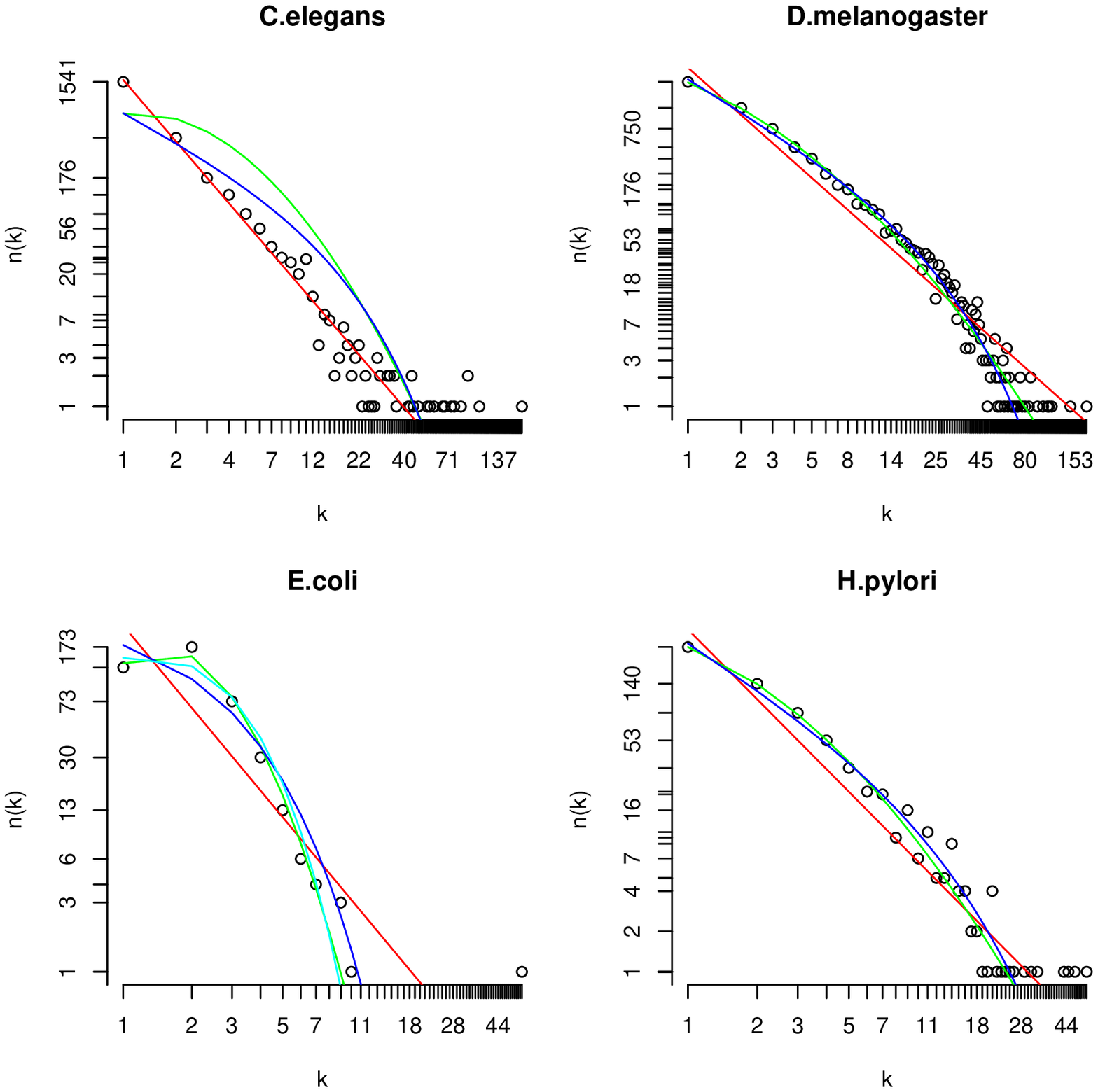,height=12cm}
\caption{Degree distributions of the protein interaction networks (o) of {\it C.elegans}, {\it D.melanogaster}, {\it E.coli} and {\it H.pylori}. The power-law (\textcolor{red}{\bf ---}), log-normal (\textcolor{green}{\bf ---}) and stretched exponential (\textcolor{blue}{\bf ---}) models are shown for all figures; for {\it E.coli} the gamma distribution (\textcolor{cyan}{\bf ---}), which performs better (measured by the Akaike weights) than either scale-free and the stretched exponential distributions.}
\end{center}
\end{figure}}\par
\par
The standard scale-free model is superior to the log-normal only for {\it C.elegans}.
The order of models (measured by decreasing Akaike weights) is M6, M5, M4, M2, M3, M1 for {\it D.melanogaster}, M6, M4, M5, M2, M3, M1 for {\it C.elegans}, M5, M6, M4, M2, M3, M1 for {\it S.cerevisiae} and {\it H.pylori}, and M5, M3, M6, M4, M2, M1 for {\it E.coli}. Thus in the light of present data the PIN degree distribution of {\it E.coli} lends more support to a Gamma distribution than to a scale-free (or even stretched scale-free) model. There is of course, no mechanistic reason why the gamma distribution should be biologically plausible but this point demonstrates that present PIN data is more complicated than predicted by simple models. Therefore  statistical model selection is needed to determine the extent to which simple models really provide insights into the intricate architecture of PINs. 
For completeness we note that model selection based on BIC results in the same ordering of models as the AIC shown here. 
\par
In table \ref{gof} we give the values of $D$ and $D^*$ for the empirical degree distributions. The estimated cumulative distribution $P(k;\hat{\theta})$ was obtained from the maximum likelihood fits of the respective models. The results in table \ref{gof} show that the maximum likelihood framework (or the respective models) sometimes cannot adequately describe the tails of the distribution ---at low and high values of the connectivity--- in some cases, where $D^*>>D$. The order of the different models suggested by $D$ for the three models generally agrees with the ordering obtained from the AIC.
  
\subsection{Analysis of MN data}
Metabolic networks aim to describe the biochemical machinery underlying cellular processes. Here we have used data from the KEGG database (\url{www.genome.jp/kegg}) with additional information from the  BRENDA database (\url{www.brenda.uni-koeln.de}). The nodes are the enzymes and an edge is added between two enzymes if one uses the products of the other as educts. 
\begin{table}[H]
\centering
\begin{tabular}{|c||c|c|c|c|c|c|}\hline
Data&{\bf M1}&{\bf M2}&{\bf M3}&{\bf M4}&{\bf M5}&{\bf M6}\\\hline
KEGG&-8276&-5247&-7676&-5946&-5054&{\bf -4178}\\
{\it H.sapiens}&-1619&-1390&-1565&-1525&-1312&{\bf -1308}\\
{\it S.cerevisiae}&-2335&-1485&-2185&-1621&-1436&{\bf -1427}\\
\hline
\end{tabular}
\vskip1mm
\label{metabtab}
\caption{Log-likelihoods for the degree distributions from table 1 applied to metabolic networks. Human and yeast data were extracted from the global database.} 
\end{table}
\par
\afterpage{\begin{figure}
\begin{center}
\epsfig{file=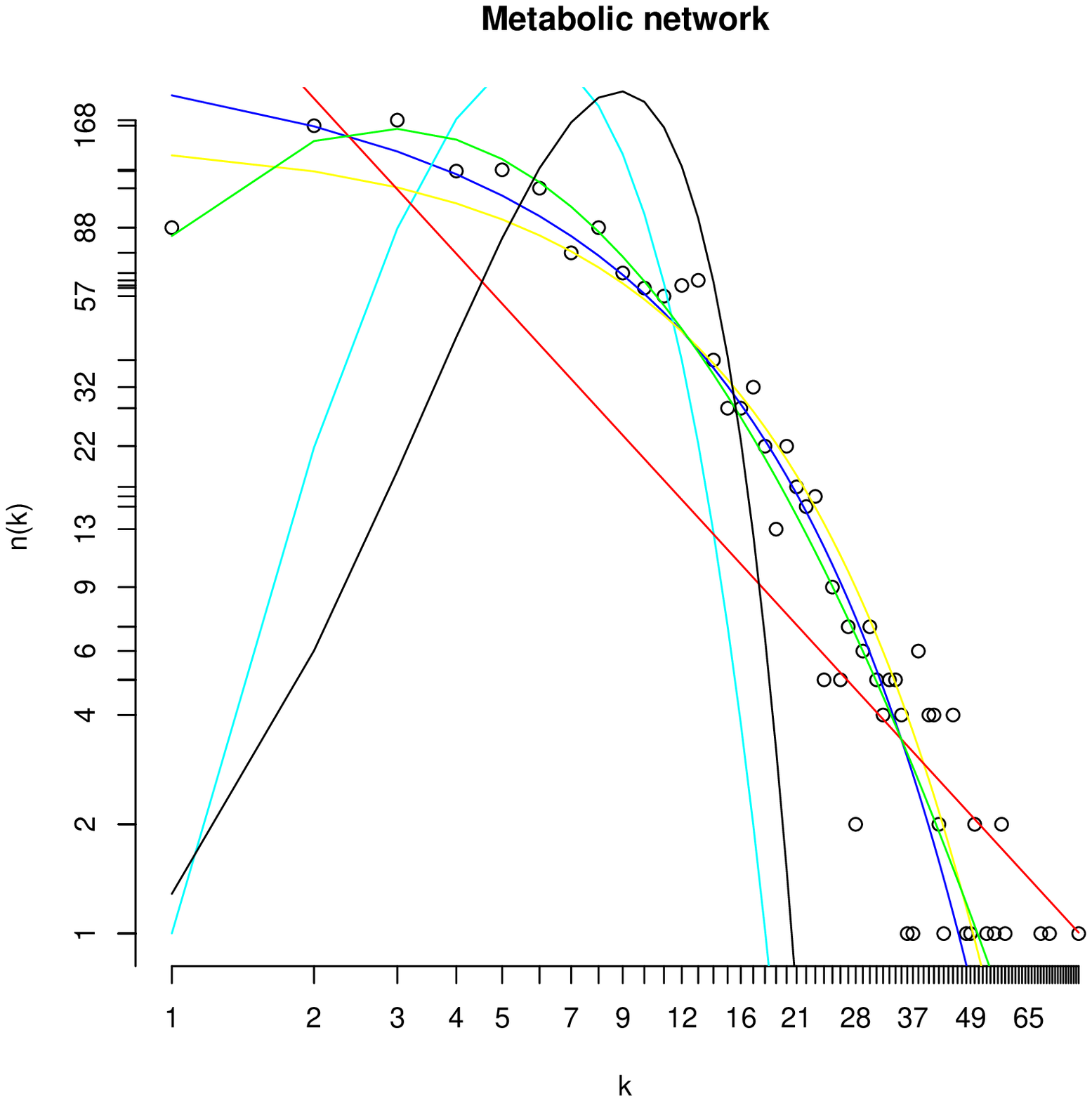,height=8cm}
\caption{Degree distributions of the metabolic network data  (o) and best-fit probability distributions: Poisson (\textcolor{black}{\bf ---}), Exponential (\textcolor{yellow}{\bf ---}), Gamma (\textcolor{cyan}{\bf ---}), Power-law (\textcolor{red}{\bf ---}), Log-normal (\textcolor{green}{\bf ---}), Stretched exponential (\textcolor{blue}{\bf ---}). The parameters of the distributions shown in this figure are the maximum likelihood estimates based on the real observed data. Ignorig low degrees ($k=1,2$) when fitting the scale-free model increases the exponent (\ie it falls off more steeply) but compared to the other models no increased performance is observed.}
\end{center}
\label{metabfig}
\end{figure}}
From figure \ref{metabfig} and table \ref{metabtab} it is apparent that the maximum likelihood scale-free model obtained from the whole network data does not provide an adequate description of the MN data. This should, however, not be too surprising as the network is only relatively small with maximum degree $k_{\max}=83$. The degree distribution appears to decay in an essentially exponential fashion but the stretched exponential has the required extra flexibility to describe the whole of the data better than the other models. Measured by goodness of fit statistics  $D$ and $D^*$, however, the log-normal model ($D=0.02$ and $D^*=0.06$) performs rather better than the stretched exponential ($D=0.07$ and $D=0.22$); the scale-free model again performs very badly ($D=1.0$ and $D^*=34.2$) for the metabolic network data.  
\par

\section{Conclusions}
We have shown that it is possible to use standard statistical methods in order to determine which probability model describes the degree distribution best. We find that the common practice of fitting a pure power-law to such experimental network data\cite{dor_book,albert_RMP2002} may obscure information contained in the degree distribution of biological networks. This is often done by identifying a range of connectivities from the log-log plots of the degree distribution which can then be fitted by a straight line. Not only is this wasteful in the sense that not all of the data is used but it may obfuscate real, especially finite-size, trends. The same will very likely hold true for other biological networks, too \cite{may_PRE2001}. The approach used here, on the other hand, (i) uses all the data, and (ii)  can be extended to assessing levels of confidence through combining a bootstrap procedure with the Akaike weights. What we have shown here, in summary, is that statistical methods can be usefully applied to protein interaction and metabolic network data. 
\par
Even though the degree distribution does not capture all (or even most) of the characteristics of real biological networks there is reason to reevaluate previous studies. We find that formally real biological networks provide very little support for the common notion that biological networks are scale-free.  Other fat-tailed probability distributions provide qualitatively and quantitatively better descriptions of the degree distributions of biological networks. For protein interaction networks we found that the log-normal and the stretched exponential offer superior descriptions of the degree distribution than the powerlaw or its finite size versions. For metabolic versions our results confirms this. Even the exponential model outperformed the scale-free model in describing the empirical degree distribution (we note that randomly growing networks are characterized by an exponentially decreasing degree distribution). The best models are all fat-tailed ---like the scale-free models--- but are not formally scale-free. Unfortunately, there is as yet no known physical model for network growth processes that would give rise to log-normal or stretched exponential degree distributions.

\par
There is thus a need to develop and study theoretical models of network growth that are better able to describe the structure of existing networks. This probably needs to be done in light of at least three constraints: (i) real networks are finite sized and thus, in the terms of statistical physics, mesoscopic systems; (ii) present network data are really only samples from much larger networks as not all proteins are included in present experimental setup (in our case {\it S.cerevisiae} has the highest fraction, 4773 out of approximately 5500-6000 proteins); the sampling properties have recently been studied and it was found that generally the degree distribution of a subnet will differ from that of the whole network. This is particularly true for scale-free networks. (iii) biological networks are under a number of functional and evolutionary constraints and proteins are more likely to interact with proteins in the same cellular compartment or those involved in the same biological process. This modularity ---and the information already availabe \eg in gene ontologies--- needs to be considered. Finally there is an additional caveat: biological networks are not static but likely to change during development. More dynamic structures may be required to deal with this type of problem.

\par
Quite generally we believe that we are now at a stage where simple models do not necessarily describe the data collected from complex processes to the extent that we would like them to. But as Burda, Diaz-Correia and Krzywicki point out \cite{burda_PRE2001}, even if a mechanistic model is not correct in detail, a corresponding statistical ensemble may nevertheless offer important insights. We believe that the statistical models employed here will also be useful in helping to identify more realistic ensembles. 
\par

The maximum likelihood, goodness of fit and other tools and methods for the analysis of network data are implemented in the NetZ R-package which is available from the corresponding author on request.
\par

{\bf Acknowledgements:} We thank the Wellcome Trust for a research fellowship (MPHS) and a research studentship (PJI). CW is supported by the Danish Cancer Society. Financial support from the Royal Society and the Carlsberg Foundation (to MPHS and CW) is also gratefully acknowledged.  We have furthermore benefitted from discussions with Eric de Silva, Bob May and Mike Sternberg.

\end{document}